\documentclass[letterpaper, 10 pt, conference]{ieeeconf}  

\IEEEoverridecommandlockouts                              

\usepackage{graphics} 
\usepackage{epsfig} 
\usepackage{mathptmx} 
\usepackage{times} 
\usepackage{mathtools,color}
\usepackage{algorithm}
\usepackage{algpseudocode}
\usepackage{amssymb}  
\usepackage{comment}
\usepackage{subcaption}
\usepackage{multirow}
\usepackage{graphicx}

\title{\LARGE \bf
Sensor Fusion Methods for Gaussian Mixture Models
}

\author{Ishan Paranjape$^{1}$, Islam Hussein$^{2}$, Jeremy Murray-Krezan$^{3}$, Sean Phillips$^{4}$, and Suman Chakravorty$^{5}$
\thanks{*Approved for public release; distribution is unlimited. Public Affairs approval \#AFRL20244853.}
\thanks{$^{1}$Ishan Paranjape is a Graduate Research Assistant with the Department of Aerospace Engineering at Texas A\&M University, College Station, TX 77843, USA
        {\tt\small ishan.paranjape@tamu.edu}}%
\thanks{$^{2}$Islam Hussein is the Director of Research \& Development with Trusted Space, Inc., Leesburg, VA 20175, USA
        {\tt\small Islam.Hussein@trustedspace.us}}%
\thanks{$^{3}$Jeremy Murray-Krezan is a Research Scientist with Trusted Space, Inc., Leesburg, VA 20175, USA
        {\tt\small Jeremy.Murray-Krezan@trustedspace.us}}%
\thanks{$^{4}$Sean Phillips is a Research Mechanical Engineer with the Space Vehicles Directorate at Air Force Research Laboratory, Kirtland, NM 87108, USA
        {\tt\small sean.phillips.9@spaceforce.mil}}%
\thanks{$^{5}$Suman Chakravorty a Professor with the Department of Aerospace Engineering, Texas A\&M University, College Station, TX 77843, USA
        {\tt\small schakrav@tamu.edu}}%
}

\begin{document}

\maketitle
\thispagestyle{empty}
\pagestyle{empty}

\begin{abstract}

Consensus is a popular technique for distributed state estimation. This formulation allows networks of connected agents or sensors to exchange information about the distribution of a set of targets with their immediate neighbors without the need of a centralized node or layer. We present decentralized consensus-based fusion techniques for a system whose target prior estimates are a weighted mixture of Gaussian probability density functions (PDFs) for the following cases: 1) in which all agents have the same \textit{a priori} Gaussian mixture estimate of the target, and 2) in which agents have different \textit{a priori} Gaussian mixture estimates of the target. For the second case, we present a formulation that fuses each agent's \textit{a priori} estimate without using local observations such that each agent's posterior estimate is the same across the network.

\end{abstract}

\section{INTRODUCTION}

Sensor fusion has become a popular tool for studying the situational awareness of various parameters, states, and dynamic systems. Sensor fusion refers to the combination of data from a set of heterogeneous or homogeneous sensors in order to reduce the uncertainty of the state of one or more dynamic systems or targets. Sensor fusion for target tracking has been important for several applications such as autonomous systems and surveillance \cite{c1,c2,c3}. Sensor fusion may be accomplished in a centralized or decentralized manner. Whereas a centralized sensor network contains a single node which fuses sensor data, a decentralized sensor network contains sensors which perform local estimates of target states based on the raw data of themselves and their neighbors. However, centralized sensor fusion has the potential for data processing bottlenecks, high communication bandwidth, and vulnerability resulting from a single point of failure. We focus on decentralized sensor fusion approaches because of their robustness, resiliency to sensor failure, and scalability. Although decentralized sensor fusion has limitations for high-level decision making, we focus our work on situations where lower-level decision making is emphasized. 

Tsitsiklis first summarized ideas and methods for decentralized sensor detection and message passing in \cite{c4}. Since then, flooding has emerged as a common decentralized information exchange scheme, along with more sophisticated decentralized information fusion methods such as Bayesian hypothesis testing and distributed Kalman filtering \cite{c5, c6, c7}. A disadvantage of flooding is that it is not scalable to large sensor networks. In Ref. \cite{c5}, \textit{Xiao et. al.} presented a method for exchanging distributed weighted least-squares (WLS) estimates in a way that converges to a global maximum-likelihood solution. The advantage that Ref. \cite{c5} shows by utilizing the information form of the Kalman Filter is that a parameter mean and covariance may be computed as a Metropolis-Hastings Markov Chain (MHMC) distributed average. Assuming jointly connected sensor networks, the distributed averaging framework may be scalable across large sensor networks.

In distributed averaging, there may exist cases in which information passed from one node or sensor to another is already conditioned upon information from the receiving node. This leads to overconfidence in the certainty of a target estimate. Following the seminal work of Refs. \cite{c8} and \cite{c9}, Ref. \cite{c10} expands upon covariance intersection (CI) as a way of placing an upper bound on the uncertainty of a state estimate. Refs. \cite{c11}, \cite{c12}, and \cite{c20} study a more specific form of CI known as iterative covariance intersection (ICI) and compare how the covariance bounds compare with distributed consensus alone, ICI, and a mixture of consensus and ICI. However, it has been noted in reviews of CI data fusion that the CI estimator will usually produce higher state uncertainties and optimal upper bounds in the uncertainty for a single time step may not produce optimal upper bounds in the uncertainty for subsequent time steps \cite{c13, c14}. For this reason, and by considering independent \textit{a priori} estimates, we ignore the ICI estimator for this article and focus solely on an MHMC-averaging consensus estimator. 

While Refs. \cite{c12} and \cite{c5} provide good frameworks for consensus estimators, both articles assume 1) linearity in dynamics and measurement models, and 2) purely Gaussian prior estimates. While a single Gaussian probability density function (PDF) is insufficient for target tracking in several cases (see Ref. \cite{c15} for an example), a Gaussian mixture model (GMM) may more accurately capture the evolution of a target's true state or truth. The objective of this research is to generalize the framework for consensus estimators to prior estimates that may be expressed as GMMs, considerably relaxing the second assumption. Furthermore, we have slightly relaxed the first assumption by utilizing a nonlinear observation or measurement model for each sensor in our network.

In this article, we present two distributed state estimation (DSE) techniques which fuse estimates expressed as Gaussian mixture models: one which fuses GMM state estimates assuming each agent has the same prior estimate and another which assumes differing prior estimates among agents. We organize the rest of our paper as follows: In Section II, we introduce a mathematical formulation for DSE using a consensus estimator. In Section III, we introduce mathematical formulations for GMM fusion for the cases of same and differing prior estimates. Finally, in Section IV, we present some examples of DSE for a simple system.



\section{DISTRIBUTED STATE ESTIMATION}\label{sec:dse}

Consider a discrete-time state-space model of a linear stochastic dynamic system with a non-linear measurement model.
\begin{equation}\label{eq1:dynModel}
    \mathbf{x}_{k+1} = \mathbf{F}_x\mathbf{x}_k + \mathbf{w}_k.
\end{equation}
\begin{equation}\label{eq2:measModel}
    \mathbf{z}_{k} = \mathbf{h}(\mathbf{x}_k) + \mathbf{v}_k.
\end{equation}

We denote the state and observation vectors as $\mathbf{x}_k \in \mathbb{R}^{n_x}$ and $\mathbf{z}_k \in \mathbb{R}^{n_z}$, respectively. Furthermore, we consider the process and observation additive noises, denoted by $\mathbf{w}_k$ and $\mathbf{v}_k$ respectively, to be zero-mean and Gaussian (i.e. $\mathbf{w}_k \sim \mathcal{N}(\mathbf{0}, \mathbf{Q})$ and $\mathbf{v}_k \sim \mathcal{N}(\mathbf{0}, \mathbf{R})$). 

DSE aims to provide a posterior PDF $p(\mathbf{x}_k|\mathbf{z}_k)$ for the dynamic system at time $t = k$, given posterior PDFs at the previous time step. In DSE, it is important to note that we do not have access to each measurement within our network topology via a centralized node. In this section, we define a framework for distributed state estimation by detailing the network topology, algorithms for consensus, and the information fusion process.

\subsection{Targets and Sensor Networks}\label{sec:netTop}

We define our sensor network as a graph data structure $\mathcal{G}(\mathcal{V}, \mathcal{E}(k))$, where $\mathcal{V} \in {v_1, v_2, ..., v_S}$ represents the $S$ sensors or nodes of our network and $\mathcal{E}(k)$ represents the set of edges or bi-directional connections between sensors at time $t = k$. Mathematically, if we denote that $(v_i, v_j) \in \mathcal{E}(k)$, then we say that nodes $v_i$ and $v_j$ communicate information bi-directionally at time $t = k$. Given a sensor $v_i$, we define the neighbors of $v_i$ as follows:
\begin{equation}\label{eq3:neighbors}
    \mathcal{N}_i(k) = \{v_i\} \cup \{v_j \in \mathcal{V} | (v_i, v_j) \in \mathcal{E}(k)\}.
\end{equation}
Furthermore, we denote the cardinality of $\mathcal{N}_i(k)$ as $|\mathcal{N}_i(k)|$. For simplicity, we also assume that each sensor within our network is homogeneous (i.e. each sensor has the same measurement model and observation noise characteristics). 

\subsection{Distributed Filtering Mathematical Preliminaries}

Since we assume zero-mean Gaussian noise in Eq. \ref{eq1:dynModel} and \ref{eq2:measModel}, as well as a nonlinear measurement model, the Extended Kalman Filter (EKF) becomes an optimal recursive filter for DSE. We denote the predicted or \textit{a priori} mean and covariance values as $\mathbf{\mu}_{x,k}^{-}$ and $\mathbf{P}_{xx,k}^{-}$, respectively, and the estimated or posterior mean and covariance values by $\mathbf{\mu}_k^{+}$ and $\mathbf{P}_{xx,k}^{+}$, respectively.

Utilizing the traditional form of the EKF for DSE would require several matrix inversion and addition steps, leading to high computational complexity for large-scale sensor networks. \cite{c16} By utilizing the \textit{information form} of the Kalman Filter, we may exchange measurement models, measurement likelihood functions, and individual posterior PDF updates. \cite{c17}\cite{c18} For both the EKF and the information form of the Kalman Filter, we linearize the measurement model given in Eq. \ref{eq2:measModel} at the predicted mean as follows:
\begin{equation}\label{eq6:measModelLin}
    \mathbf{H}_{x,k} = \frac{\partial h(\mathbf{x}_k)}{\partial \mathbf{x}_k}\bigg|_{\mathbf{\mu_{x,k}^{-}}}.
\end{equation}

We define the two main variables of the information Kalman Filter, the information matrix and the information vector (denoted as $\mathbf{Y}_k$ and $\mathbf{y}_k$, respectively) as
\begin{equation}\label{eq7:infoP}
    \mathbf{Y}_{k} = (\mathbf{P}_{xx,k})^{-1}
\end{equation}
\begin{equation}\label{eq8:infomu}
    \mathbf{y}_{k} = (\mathbf{P}_{xx,k})^{-1}\mathbf{x}_k = \mathbf{Y}_{k}\mathbf{x}_k.
\end{equation}

We may then utilize the following equations to recursively compose the prediction step of our information form Kalman Filter.
\begin{equation}\label{eq9:infMat}
    \mathbf{Y}_k^{-} = [\mathbf{F}_x (\mathbf{Y}_{k-1}^{+})^{-1} \mathbf{F}_x^{T} + \mathbf{Q}]^{-1}.
\end{equation}
\begin{equation}\label{eq10:infVec}
    \mathbf{y}_k^{-} = \mathbf{Y}_k^{-} \mathbf{F}_x \mathbf{Y}_{k-1}^{+} \mathbf{y}_{k-1}^{+}.
\end{equation}

For a centralized filter, each agent that makes an observation of the target would transmit information to a central node or aggregator. The aggregator would then utilize Eq. \ref{eq7:infoP} - \ref{eq10:infVec} to form its predicted estimate of the target state. Each node $j \in {1, 2, ..., N}$ within the network would then transmit its own information vector and information matrix in the form of
\begin{equation}\label{eq11:nodeVec}
    \delta \mathbf{i}_{j,k} = \mathbf{H}_{j,k}^{T} \mathbf{R}_j^{-1} \mathbf{z}_{j,k}.
\end{equation}
\begin{equation}\label{eq12:nodeMat}
    \delta \mathbf{I}_{j,k} = \mathbf{H}_{j,k}^{T} \mathbf{R}_j^{-1} \mathbf{H}_{j,k}.
\end{equation}
In Eq. \ref{eq11:nodeVec}, $\mathbf{z}_{j,k}$ refers to the observation made by Agent $j$ at time step $k$, $\mathbf{R}_j$ refers to the observation noise matrix, and $\mathbf{H}_{j,k}$ represents the linearized measurement model explained in Eq. \ref{eq6:measModelLin}. As the centralized aggregator draws $\delta \mathbf{i}_{j,k}$ and $\delta \mathbf{I}_{j,k}$ from each agent, it performs an aggregate posterior estimate in the form of
\begin{equation}\label{eq13:postVec}
    \mathbf{y}_k^{+} = \mathbf{y}_k^{-} + \sum_{j = 1}^{N} \delta \mathbf{i}_{j,k}. 
\end{equation}
\begin{equation}\label{eq14:postMat}
    \mathbf{Y}_k^{+} = \mathbf{Y}_k^{-} + \sum_{j = 1}^{N} \delta \mathbf{I}_{j,k}. 
\end{equation}
However, for DSE, no such centralized node exists. For this reason, we look to more sophisticated methods of state estimation. In the following subsection, we describe the utility of consensus-based estimation.

\subsection{Consensus-Based Estimator}

In order to perform distributed averaging using consensus-based estimation, one must be able to compute $\delta \mathbf{i}$ and $\delta \mathbf{I}$ from Eq. \ref{eq11:nodeVec} and \ref{eq12:nodeMat} for all time steps $k$. We express the updates in terms of the averages across our network's agents as follows:
\begin{equation}\label{eq15:postVec}
    \delta \mathbf{i}_k = S * \{\frac{1}{S} \sum_{j=1}^{S} \delta \mathbf{i}_{j,k}\}
\end{equation}
\begin{equation}\label{eq16:postMat}
    \delta \mathbf{I}_k = S * \{\frac{1}{S} \sum_{j=1}^{S} \delta \mathbf{I}_{j,k}\}
\end{equation}
We show \textit{\{\}} in the equations above to emphasize that the expression within the \textit{\{\}} will be the final result of consensus-based averaging for each agent. If each agent knows the size of the (connected) network and each sensor within the network has the same \textit{a priori} information, then each agent can perform its own update without the need of a centralized aggregator. 

We utilize the distributed averaging method from \cite{c5} due to its minimal number of assumptions about network topology and reliance of network exchange between neighboring nodes. Assuming the network is connected (i.e. each node within the network is directly or indirectly reachable to all other nodes within that network), this DSE method will enable us to reach a consensus value for all agents. 

In particular, the aforementioned DSE method relies upon the Metropolis-Hastings Markov Chain (MHMC), a linear, iterative consensus filter. With MHMC, each node assigns a weight to its own $\delta \mathbf{i}$ or $\delta \mathbf{I}$, as well as those of its neighbors. MHMC weights between two nodes $i$ and $j$, $\gamma_{i,j}$, are computed as follows:
\begin{equation}\label{eq17:mhmcWeights}
    \gamma_{i,j}^{(l)} = \begin{cases} 
      \frac{1}{max(|\mathcal{N}_{i}^{(l)}|, |\mathcal{N}_{j}^{(l)}|)} & if (i, j) \in \mathcal{E}^{(l)} \\
      1 - \sum_{\{i,m\} \in \mathcal{E}^{(l)}} \gamma_{i,m}^{(l)} & if 
      i = j \\
      0 & otherwise. 
   \end{cases}
\end{equation}
In the piecewise equation above, $(l)$ represents the iteration number, and we make use of notation defined in Section \ref{sec:netTop}. Furthermore, each iteration of MHMC occurs at a timescale that is much lower than the time gap between two consecutive observations. With MHMC weights calculated, we may iterate to converged values for $\delta \mathbf{i}$ and $\delta \mathbf{I}$ as follows:
\begin{equation}\label{eq18:informationVecIteration}
    \delta \mathbf{i}^{(l+1)} = \sum_{j=1}^{|\mathcal{N}^{(l)}|} \gamma_{i,j}^{(l)} \delta \mathbf{i}^{(l)}
\end{equation}
\begin{equation}\label{eq19:informationMatIteration}
    \delta \mathbf{I}^{(l+1)} = \sum_{j=1}^{|\mathcal{N}^{(l)}|} \gamma_{i,j}^{(l)} \delta \mathbf{I}^{(l)}
\end{equation}
Although we assume in this framework that the prior estimate of each agent in the network is the same and that our prior estimates may be expressed as purely Gaussian PDFs, prior estimates in practice will be different due to differing levels of network disconnection and may have to be approximated using several weighted Gaussian PDFs. In the proceeding section, we explore two sensor fusion methods: 1) in which all agents' prior estimates are expressed as a mixture of Gaussian PDFs, and 2) in which all agents' prior estimates are different and may be expressed as a mixture of Gaussian PDFs.


\section{GAUSSIAN MIXTURE MODEL FUSION}\label{sec:GMMfuse}

Consider a decentralized network of $S$ agents. The \textit{a priori} PDF of each agent is expressed as a Gaussian Mixture Model (GMM). Mathematically, we express these GMMs as follows:
\begin{subequations}\label{eq20:GMMcomps}
    \begin{align}
        p(\mathbf{x}_1) &= \sum_{i=1}^{N_1} \omega_{1,i} \mathcal{N}(\mathbf{\mu}_{x,1,i},P_{xx,1,i}) \\
        p(\mathbf{x}_2) &= \sum_{i=1}^{N_2} \omega_{2,i} \mathcal{N}(\mathbf{\mu}_{x,2,i},P_{xx,2,i}) \\
        &\vdotswithin{=} \notag\\
        p(\mathbf{x}_S) &= \sum_{i=1}^{N_S} \omega_{S,i} \mathcal{N}(\mathbf{\mu}_{x,S,i},P_{xx,S,i})
    \end{align}
\end{subequations}
In the equations above, we omit the subscript representing the timestep because we only consider a single step fusion process for this work. We assume that our network contains $S$ agents, and that each agent has \textit{a priori} estimates with Gaussian component PDFs with weights $\omega$. In this section, we describe our GMM fusion process for two different cases: 1) in which all \textit{a priori} estimates across all agents within the network are the same, and 2) in which the \textit{a priori} estimates are different.

\subsection{Homogeneous Prior Estimates}\label{sec3A:homEst}

We shall first consider the case in which all agents are connected before receiving measurements. Mathematically, from Eq. \ref{eq20:GMMcomps}, this means that $p(\mathbf{x}_1) = p(\mathbf{x}_2) = ... = p(\mathbf{x}_S) = p(\mathbf{x})$, $N_1 = N_2 = ... = N_S = N$, $\omega_{1,i} = \omega_{2,i} = ... = \omega_{S,i} = \omega_{i}$, $\mu_{x,1,i} = \mu_{x,2,i} = ... = \mu_{x,S,i} = \mu_{x,i}$, and $P_{xx,1,i} = P_{xx,2,i} = ... = P_{xx,S,i} = P_{xx,i}$. In order to develop our framework for fusion, we require DSE for all component mean vectors, component covariance matrices, and component weights. 

We begin deriving our posterior PDF by assuming each agent receives an observation. Substituting our \textit{a priori} estimate $p(\mathbf{x})$ with its definition from Eq. \ref{eq20:GMMcomps}, we may use Bayes' rule to derive our posterior PDF as follows. 
\begin{align} \label{eq21:postPDFderiv1} \begin{split}
    p^{+} (\mathbf{x} | \mathbf{z}_1, \mathbf{z}_1, \cdots, \mathbf{z}_S) =  \frac{p(\mathbf{x}) p(\mathbf{z}_1, \mathbf{z}_2, \cdots, \mathbf{z}_S|\mathbf{x})}{p(\mathbf{z}_1, \mathbf{z}_2, \cdots, \mathbf{z}_S)} \\
    = \frac{p(\mathbf{x}) p(\mathbf{z}_1|\mathbf{x}) p(\mathbf{z}_2|\mathbf{x}) \cdots p(\mathbf{z}_S|\mathbf{x})}{\int p(\mathbf{\zeta}) p(\mathbf{z}_1|\mathbf{\zeta}) p(\mathbf{z}_2|\mathbf{\zeta}) \cdots p(\mathbf{z}_S|\mathbf{\zeta}) d\mathbf{\zeta}} \\
    = \frac{[\sum_{i=1}^{N} \omega_i \mathcal{N}(\mathbf{\mu}_{x,i},P_{xx,i})][\prod_{s=1}^{S} p(\mathbf{z}_s|\mathbf{x})]}{\int [\sum_{j=1}^{N} \omega_j \mathcal{N}(\mathbf{\mu}_{x,j},P_{xx,j})][\prod_{s=1}^{S} p(\mathbf{z}_s|\mathbf{\zeta})] d\mathbf{\zeta}} \\
    = \frac{[\sum_{i=1}^{N} \omega_i \mathcal{N}(\mathbf{\mu}_{x,i},P_{xx,i})][\prod_{s=1}^{S} p(\mathbf{z}_s|\mathbf{x})]}{\sum_{j=1}^{N} \omega_j \int \mathcal{N}(\mathbf{\mu}_{x,j},P_{xx,j}) \prod_{s=1}^{S} p(\mathbf{z}_s|\mathbf{\zeta}) d\mathbf{\zeta}}.
\end{split}\end{align}
We assume that each agent's measurement likelihood function $p(\mathbf{z}_s|\mathbf{x})$ for $s = 1, 2,\cdots, S$ is independent of one another such that we can expand $p(\mathbf{z}_1, \mathbf{z}_2, \cdots, \mathbf{z}_S|\mathbf{x})$ into the product above. The denominator of Eq. \ref{eq21:postPDFderiv1} is expanded into an integral per the continuous form of the law of total probability. Now that we have expanded the fraction above, we will be able to express the posterior PDF as a product of the component weight update and the component PDF update, the latter of which is a Gaussian PDF as a function of the posterior mean and covariance. 

We first simplify the denominator of Eq. \ref{eq21:postPDFderiv1}. We define a likelihood function $l_i(\mathbf{z}_{1 \cdots S})$, which is the probability that all agents' observations came from their prior estimate's $i$-th GMM component, as the integral within this denominator.
\begin{equation} \label{eq22:likelihood}
    l_i(\mathbf{z}_{1 \cdots S}) \equiv \int [\prod_{s=1}^{S} p(\mathbf{z}_s|\mathbf{\chi})] \mathcal{N}(\mathbf{\mu}_{x,i}, \mathbf{P}_{xx,i}) d\mathbf{\chi}.
\end{equation}
We use the above definition to simplify and rearrange our posterior PDF equation.
\begin{align} \label{eq23:postPDFderiv1} \begin{split}
    p^{+} (\mathbf{x} | \mathbf{z}_1, \mathbf{z}_1, \cdots, \mathbf{z}_S) = \sum_{i=1}^{N} \frac{\omega_i \mathcal{N}(\mathbf{\mu}_{x,i},P_{xx,i})\prod_{s=1}^{S} p(\mathbf{z}_s|\mathbf{x})}{\sum_{j=1}^{N} \omega_{j} l_j(\mathbf{z}_{1 \cdots S})} \\
    = \sum_{i=1}^{N} \frac{\omega_i \mathcal{N}(\mathbf{\mu}_{x,i},P_{xx,i})\prod_{s=1}^{S} p(\mathbf{z}_s|\mathbf{x})}{\sum_{j=1}^{N} \omega_{j} l_j(\mathbf{z}_{1 \cdots S})} \space \frac{l_i(\mathbf{z}_{1 \cdots S})}{l_i(\mathbf{z}_{1 \cdots S})} \\
    = \sum_{i=1}^{N} \frac{\omega_i l_i(z_{1 \cdots S})}{\sum_{j=1}^{N} \omega_{j} l_j(\mathbf{z}_{1 \cdots S})} \space \frac{[\prod_{s=1}^{S} p(\mathbf{z}_s|\mathbf{x})]\mathcal{N}(\mathbf{\mu}_{x,i},P_{xx,i})}{l_i(\mathbf{z}_{1 \cdots S})} \\
    = \sum_{i=1}^{N} \omega_i^{+} p_i^{+}(\mathbf{x}|\mathbf{z}_1, \mathbf{z}_2, \cdots, \mathbf{z}_S).
\end{split}\end{align}
From Eq. \ref{eq23:postPDFderiv1}, we see that the posterior PDF, like the prior PDF, is a GMM, with a discrete update and continuous update given by Eqs. \ref{eq24a:dUpdate} and \ref{eq24b:cUpdate}, respectively.
\begin{subequations} \label{eq24:cdUpdates}
    \begin{align}
        \omega_i^{+} = \frac{\omega_i l_i(z_{1 \cdots S})}{\sum_{j=1}^{N} \omega_{j} l_j(\mathbf{z}_{1 \cdots S})} \label{eq24a:dUpdate} \\
        p_i^{+}(\mathbf{x}|\mathbf{z}_1, \mathbf{z}_2, \cdots, \mathbf{z}_S) = \frac{\prod_{s=1}^{S} p(\mathbf{z}_s|\mathbf{x}) \mathcal{N}(\mathbf{\mu}_{x,i},P_{xx,i})}{l_i(\mathbf{z}_{1 \cdots S})} \label{eq24b:cUpdate}
    \end{align}
\end{subequations}

In Section \ref{sec:dse}, we show how to fuse a single-component Gaussian PDF across a decentralized network. Recognizing that the continuous update in Eq. \ref{eq24b:cUpdate} may be performed by the consensus estimation framework described in Section \ref{sec:dse}, we are able to find all component posterior means and covariance matrices. However, finding the discrete (i.e. weight) update in Eq. \ref{eq24a:dUpdate} requires some mathematical manipulation, notably, the calculation of the likelihood term $l_i(\mathbf{z}_{1 \cdots S})$.

There are two methods of calculating $l_i(\mathbf{z}_{1 \cdots S})$, the first of which was introduced in Ref. \cite{c19}. With this method, one can see that 
\begin{equation}\label{eq25:priorOverPost}
    l_i(\mathbf{z}_{\cdots S}) = \frac{p_i(\mathbf{x}) \prod_{s=1}^{S} p(\mathbf{z}_s|\mathbf{x})}{p_i^{+} (\mathbf{x}|\mathbf{z}_1, \mathbf{z}_2, \cdots, \mathbf{z}_S)}.
\end{equation}
Per our definition from Eq. \ref{eq22:likelihood}, it is clear that the likelihood function is \textit{not} a function of $\mathbf{x}$. As a result, Ref. \cite{c19} implies that we may choose some random state vector $\mathbf{x}_c$ such that
\begin{equation}\label{eq26:constantCase}
    l_i(\mathbf{z}_{1 \cdots S}) = \frac{p_i(\mathbf{x}_c) \prod_{s=1}^{S} p(\mathbf{z}_s|\mathbf{x}_c)}{p_i^{+} (\mathbf{x}_c|\mathbf{z}_1, \mathbf{z}_2, \cdots, \mathbf{z}_S)},
\end{equation}
which is possible to calculate with the prior and posterior PDFs that we obtained with our continuous update. With the above results, we can calculate the weight update for each component $i$:
\begin{align} \label{eq27:weightUpdate} \begin{split}
    \omega_i^{+} = \frac{\omega_i \frac{p_i(\mathbf{x}_c) \prod_{s=1}^{S} p(\mathbf{z}_s|\mathbf{x}_c)}{p_j^{+} (\mathbf{x}_c|\mathbf{z}_1, \mathbf{z}_2, \cdots, \mathbf{z}_S)}}{\sum_{j=1}^{N} \omega_j \frac{p_i(\mathbf{x}_c) \prod_{s=1}^{S} p(\mathbf{z}_s|\mathbf{x}_c)}{p_j^{+} (\mathbf{x}_c|\mathbf{z}_1, \mathbf{z}_2, \cdots, \mathbf{z}_S)}} \\
    = \frac{\omega_i \frac{p_i(\mathbf{x}_c)}{p_i^{+} (\mathbf{x}_c|\mathbf{z}_1, \mathbf{z}_2, \cdots, \mathbf{z}_S)}}{\sum_{j=1}^{N} \omega_j \frac{p_j(\mathbf{x}_c)}{p_j^{+} (\mathbf{x}_c|\mathbf{z}_1, \mathbf{z}_2, \cdots, \mathbf{z}_S)}}
\end{split}\end{align}

One disadvantage of this method is that it requires a good choice of $\mathbf{x}_c$ that works for \textit{all} GMM components. If $\mathbf{x}_c$ lies too many $\sigma$-intervals outside of some GMM component, then the formula above is ill-conditioned. To address this ill-conditioning, we present the second method. 

We start by noting that one constant we can choose for $\mathbf{x}_c$ is $\mathbf{\mu}_{x,i}^{-}$, or the prior mean of the $i$-th GMM component, such that
\begin{align} \label{eq28:meanPlugin} \begin{split}
    l_i(\mathbf{z}_{1 \cdots S}) = \frac{p_i(\mathbf{x})}{p_i^{+}(\mathbf{x} | \mathbf{z}_{1 \cdots S})} \prod_{s=1}^{S} p_i(\mathbf{z}_s|\mathbf{x}) \\
    = \frac{p_i(\mathbf{\mu}_{x,i}^{-})}{p_i^{+}(\mathbf{\mu}_{x,i}^{-} | \mathbf{z}_{1 \cdots S})} \prod_{s=1}^{S} p_i(\mathbf{z}_s|\mathbf{\mu}_{x,i}^{-}).
\end{split}\end{align}
Although this choice of $\mathbf{x}_c$ mitigates ill-conditioning since the prior-to-posterior density ratio will be relatively higher, we will not be able to access the product term $\prod_{s=1}^{S} p(\mathbf{z}_s|\mathbf{\mu}_{x,i}^{-})$ directly with a decentralized sensor network. We can, however, convert this product term so that we can evaluate a representative term using the consensus technique outlined in Section \ref{sec:dse}. As demonstrated in Ref. \cite{c20}, we start by taking the natural logarithm of the product term, noting that the product of terms within a logarithmic function is the sum of the logarithm of those terms. 
\begin{equation} \label{eq29:dseLambda}
    \ln{[\prod_{s=1}^{S} p_i(\mathbf{z}_s | \mathbf{\mu}_{x,i}^{-})]} = \sum_{s=1}^{S} \ln{[p_i(\mathbf{z}_s | \mathbf{\mu}_{x,i}^{-})]}. 
\end{equation}
Since the term on the right hand side of Eq. \ref{eq29:dseLambda} is a sum over the network, and that each agent $s$ has access to its own observation and all of its \textit{a priori} GMM component means, we can introduce a quantity $\Tilde{l}_i$ which represents the logarithm of this measurement likelihood product for some \textit{a priori} GMM component $i$. We show the $\{\}$ terms to draw an analogy between this equation and Eqs. \ref{eq15:postVec} and \ref{eq16:postMat}, namely that one can use consensus to calculate the logarithm of the measurement likelihood product.
\begin{equation} \begin{split} \label{eq30edit:deltaLambda}
    \Tilde{l}_i = \sum_{s=1}^{S} \ln{[p_i(\mathbf{z}_s | \mathbf{\mu}_{x,i}^{-})]} \\
    = S * \{\frac{1}{S} \sum_{s=1}^{S} \ln{[p_i(\mathbf{z}_n | \mathbf{\mu}_{x,i}^{-})]}\}
\end{split}\end{equation}


$\Tilde{l}_i$ represents the natural logarithm of the measurement likelihood product. This may be generalized across all GMM components of the \textit{a priori} estimate. Referencing Eq. \ref{eq28:meanPlugin}, we notice that we can calculate the likelihood that all observations $\mathbf{z}_1, \mathbf{z}_2, \cdots, \mathbf{z}_S$ come from the $i$-th prior GMM component. Referencing Eq. \ref{eq24:cdUpdates}, we can verify that we have all of the tools to perform fusion such that all posterior GMM components are also the same throughout the connected network. We organize our framework into Algorithm \ref{alg:cap}. In the next subsection, we shall show a fusion method when GMM prior estimates between agents are different.

\begin{algorithm}
\caption{Consensus on Homogeneous \textit{A Priori} GMMs}\label{alg:cap}
\begin{algorithmic}[1]
    \State \parbox[t]{\dimexpr\linewidth-\algorithmicindent-1.5em}{\textbf{Input}: $[\mathbf{y}_{i,s} (t_0), \mathbf{Y}_{i,s} (t_0)]$ for all GMM components $i \in \{1, 2, \cdots, N\}$ and all agents/sensors $s \in \{1, 2, \cdots, S\}$}
    \For{$i \in \{1, 2, \cdots, N\}$}
        \For{$s \in \{1, 2, \cdots, S\}$}
            \State \parbox[t]{\dimexpr\linewidth-\algorithmicindent-1.5em}{Use Eqs. \ref{eq7:infoP}-\ref{eq10:infVec} to calculate prior estimates for the information vector and information matrix $[\mathbf{y}_{i,s}^{-} (t_1), \mathbf{Y}_{i,s}^{-} (t_1)]$ given $[\mathbf{y}_{i,s}^{+} (t_0), \mathbf{Y}_{i,s}^{+} (t_0)]$.}
            \State \parbox[t]{\dimexpr\linewidth-\algorithmicindent-1.5em}{Utilize the agent's local observation $\mathbf{z}_s(t_1)$ and compute the measurement noise covariance and measurement Jacobian per Eq. \ref{eq6:measModelLin} to obtain $[\mathbf{H}_{i,s}(t_1), \mathbf{R}_{i,s}(t_1)]$.}
            \State Calculate the agent's local information update.
            \State $\begin{aligned}
            \delta \mathbf{i}_{i,s}(t_1) &= \mathbf{H}_{i,s}^{T}(t_1) \mathbf{R}^{-1}_{i,s}(t_1) \mathbf{z}_s(t_1) \\
            \delta \mathbf{I}_{i,s}(t_1) &= \mathbf{H}_{i,s}^{T}(t_1) \mathbf{R}^{-1}_{i,s}(t_1) \mathbf{H}_{i,s}(t_1) \\
            \Tilde{l}_{i,s}(t_1) &= 0
            \end{aligned}$ 
            \State \parbox[t]{\dimexpr\linewidth-\algorithmicindent-1.5em} {Initialize the iteration variable for consensus (i.e. $l = 0$).}
            \State \parbox[t]{\dimexpr\linewidth-\algorithmicindent-1.5em}{Set $\overline{\delta \mathbf{i}}_{i,s}^{(l=0)} = \delta \mathbf{i}_{i,s}(t_1)$, $\overline{\delta \mathbf{I}}_{i,s}^{(l=0)} = \delta \mathbf{I}_{i,s}(t_1)$, and $\overline{\Tilde{l}}_{i,s}^{(l=0)} = \Tilde{l}_{i,s}(t_1)$.} 
            \While{\textit{NOT CONVERGED}}
                \State BROADCAST $[\overline{\delta \mathbf{i}}_{i,s}^{(l)}, \overline{\delta \mathbf{I}}_{i,s}^{(l)}, \overline{\Tilde{l}}_{i,s}^{(l)}]$.
                \State \parbox[t]{\dimexpr\linewidth-\algorithmicindent-1.5em}{RECEIVE $[\overline{\delta \mathbf{i}}_{i,s'}^{(l)}, \overline{\delta \mathbf{I}}_{i,s'}^{(l)}, \overline{\Tilde{l}}_{i,s'}^{(l)}]$ $\forall s' \in  \mathcal{N}_s^{(l)}(t_1)$.}
                \State \parbox[t]{\dimexpr\linewidth-\algorithmicindent-2.75em}{Perform a single iteration of MHMC as described by Ref. \cite{c5} to obtain $[\overline{\delta \mathbf{i}}_{i,s}^{(l+1)}, \overline{\delta \mathbf{I}}_{i,s}^{(l+1)}, \overline{\Tilde{l}}_{i,s}^{(l+1)}]$.}
                \State $l \gets l + 1$
            \EndWhile
            \State \parbox[t]{\dimexpr\linewidth-\algorithmicindent-1.5em} {Calculate posterior estimates with $\mathbf{y}_{i,s}^{+}(t_1) = \mathbf{y}_{i,s}^{-}(t_1) + S*\overline{\delta \mathbf{i}}_{i,s}^{(l)}$ and $\mathbf{Y}_{i,s}^{+}(t_1) = \mathbf{Y}_{i,s}^{-}(t_1) - S*\overline{\delta \mathbf{I}}_{i,s}^{(l)}$.} 
            \State \parbox[t]{\dimexpr\linewidth-\algorithmicindent-1.5em} {Calculate measurement likelihood $\Tilde{l}_{i,s}$ for the weight updates with}
            \State $\Tilde{l}_{i,s} = S*\overline{\Tilde{l}}_{i,s}^{(l)}$
        \EndFor
    \EndFor
\end{algorithmic}\end{algorithm}

\subsection{Nonhomogeneous Prior Estimates}

In general, each agent within a network will not have the same \textit{a priori} estimate of the target state as the others, such as when network disconnections occur. Before computing a distributed fused estimate using the observations $\mathbf{z}_1, \mathbf{z}_2, \cdots, \mathbf{z}_S$, it is a good idea to find a way to fuse prior estimates such that each agent within the network has an intermediate, homogeneous \textit{a priori} estimate prior to the Bayesian update process.  For simplicity, let us assume that there are only two connected agents within our decentralized network, whose prior PDFs are expressed by Eq. \ref{eq20:GMMcomps} and are independent. The estimates of two agents will be independent if they have been disconnected for sufficiently long and have access only to disjoint sets of observations. In the following, we show how such GMM estimates can be fused utilizing the means of one of the GMM estimates as the observations for the components of the other GMM estimate.

To fuse these two PDFs' prior estimates using our means-as-observations framework, we first have to recognize that any of the component means of Agent 1's \textit{a priori} estimate can serve as a full state estimate of the target truth for any of Agent 2's component means, similar to how an observation reflects the full or partial state of a target through a linear or nonlinear mapping, and vice versa. Abusing notation slightly, let us assume that the \textit{a priori} covariances for some GMM component $i$ and some GMM component $j$ for Agents 1 and 2, respectively, are $\mathbf{P}_1$ and $\mathbf{P}_2$, and their \textit{a priori} means are $\mathbf{\mu}_1$ and $\mathbf{\mu}_2$. 
Just as we express system measurements with Eq. \ref{eq2:measModel}, we say that 
\begin{subequations} \label{eq.32:measAnalog} \begin{align}
    \mathbf{\mu}_1 = h(\mathbf{x}_2) +  \delta \mathbf{x}_1 = \mathbf{x}_2 + \delta \mathbf{x}_1 \label{eq32a:x1}\\
    \mathbf{\mu}_2 = h(\mathbf{x}_1) +  \delta \mathbf{x}_2 =\mathbf{x}_1 + \delta \mathbf{x}_2. \label{eq32b:x2}
\end{align}\end{subequations}
Here, $h(.)$ is simply a linear transformation expressed as an identity matrix since $\mathbf{x}_1$ and $\mathbf{x}_2$ have the same size and units as each other. $\delta \mathbf{x}_1$ and $\delta \mathbf{x}_2$ represent noise random variable vectors which we assume to be independent of the state vectors to which they are added. We can make this assumption because, after sufficiently long network disconnections, the random variable vector pairs $(\mathbf{x}_1, \delta \mathbf{x}_2)$ and $(\mathbf{x}_2, \delta \mathbf{x}_1)$ will become uncorrelated. 

If we assume that $p(\mathbf{x}_1)$ and $p(\mathbf{x}_2)$ are Gaussian PDFs representing some \textit{a priori} estimate component of Agents 1 and 2, respectively, then we may express the Bayesian update for that particular component association between Agents 1 and 2 as follows:
\begin{subequations} \label{eq33edit:bayesians}
    \begin{align}
        p^{+}(\mathbf{x}_1|\mathbf{\mu}_2) = \frac{p(\mathbf{x}_1) p(\mathbf{\mu}_2|\mathbf{x}_1)}{p(\mathbf{\mu}_2)} \label{eq33editA:a1} \\
        p^{+}(\mathbf{x}_2|\mathbf{\mu}_1) = \frac{p(\mathbf{x}_2) p(\mathbf{\mu}_1|\mathbf{x}_2)}{p(\mathbf{\mu}_1)} \label{eq33editB:a2}.
    \end{align}
\end{subequations}

Within the denominators of Eqs. \ref{eq33editA:a1} and \ref{eq33editB:a2}, we may notice that $\mathbf{\mu}_1$ and $\mathbf{\mu}_2$ are linear transformations of $\mathbf{x}_2$ and $\mathbf{x}_1$, respectively. Since the random variable vector pairs $(\mathbf{x}_1, \delta \mathbf{x}_2)$ and $(\mathbf{x}_2, \delta \mathbf{x}_1)$ are assumed independent and Gaussian, we are able to observe that $p(\mathbf{\mu}_1)$ and $p(\mathbf{\mu}_2)$, the likelihoods of the component means, are the same.
\begin{equation}\label{eq33:pfOfSymmetry}\begin{split}
    p(\mathbf{\mu}_1) = p_g(\mathbf{\mu}_1; \mathbf{\mu}_2, \mathbf{P}_{xx,1} + \mathbf{P}_{xx,2}) \\
    = p_g(\mathbf{\mu}_1 - \mathbf{\mu}_2; \mathbf{0}, \mathbf{P}_{xx,1} + \mathbf{P}_{xx,2}) \\
    = p_g(\mathbf{\mu}_2 - \mathbf{\mu}_1; \mathbf{0}, \mathbf{P}_{xx,1} + \mathbf{P}_{xx,2}) \\
    = p_g(\mathbf{\mu}_2; \mathbf{\mu}_1, \mathbf{P}_{xx,1} + \mathbf{P}_{xx,2}) = p(\mathbf{\mu}_2)
\end{split}\end{equation}
The function $p_g$ represents a Gaussian PDF, and we use this form instead of our usual $\mathcal{N}$ to show symmetry between the distributions of $\mathbf{\mu}_1$ and $\mathbf{\mu}_2$. This result will be important for the weight update. We can now show that the fused \textit{a priori} means and fused \textit{a priori} covariances are also the same. 
We first focus on the fused covariance. Just as in Section \ref{sec3A:homEst}, a Kalman update is performed. With our above assumptions, the fused covariances for each agent may be expressed as follows:
\begin{subequations} \label{eq34:fusedCovs}
    \begin{align}
        \mathbf{P}_1^{+} = \mathbf{P}_1 -\mathbf{P}_1(\mathbf{P}_1 + \mathbf{P}_2)^{-1}\mathbf{P}_1. \\
        \mathbf{P}_2^{+} = \mathbf{P}_2 -\mathbf{P}_2(\mathbf{P}_1 + \mathbf{P}_2)^{-1}\mathbf{P}_2.
    \end{align}
\end{subequations}
To prove that $\mathbf{P}_1^{+} = \mathbf{P}_2^{+}$, we use the Woodbury matrix identity as follows:
\begin{equation}\label{eq35:pfOfFusedCov}\begin{split}
    \mathbf{P}_1^{+} = \mathbf{P}_1 -\mathbf{P}_1(\mathbf{P}_1 + \mathbf{P}_2)^{-1}\mathbf{P}_1 \\ 
    = [(\mathbf{P}_1)^{-1}]^{-1} - [(\mathbf{P}_1)^{-1}]^{-1} \mathbf{I} \{[(\mathbf{P}_2)^{-1}]^{-1} \\
    + \mathbf{I} [(\mathbf{P}_1)^{-1}]^{-1} \mathbf{I}\} \mathbf{I} [(\mathbf{P}_1)^{-1}]^{-1} \\
    = [(\mathbf{P}_1)^{-1} + (\mathbf{P}_2)^{-1}]^{-1} \\
    = [(\mathbf{P}_2)^{-1} + (\mathbf{P}_1)^{-1}]^{-1} \\
    = [(\mathbf{P}_2)^{-1}]^{-1} - [(\mathbf{P}_2)^{-1}]^{-1} \mathbf{I} \{[(\mathbf{P}_1)^{-1}]^{-1} \\
    + \mathbf{I} [(\mathbf{P}_2)^{-1}]^{-1} \mathbf{I}\} \mathbf{I} [(\mathbf{P}_2)^{-1}]^{-1} \\
    = \mathbf{P}_2 -\mathbf{P}_2(\mathbf{P}_1 + \mathbf{P}_2)^{-1}\mathbf{P}_2 = \mathbf{P}_2^{+}.
\end{split}\end{equation}

We shall utilize the results from the derivation in Eq. \ref{eq35:pfOfFusedCov} to show that the fused means of both agents are also the same. A Kalman update for the means with our above assumptions will yield the following values:
\begin{subequations} \label{eq36:fusedMeans}
    \begin{align}
        \mathbf{\mu}_1^{+} = \mathbf{\mu}_1 + \mathbf{P}_1 (\mathbf{P}_1 + \mathbf{P}_2)^{-1} (\mathbf{\mu}_2 - \mathbf{\mu}_1)\\
        \mathbf{\mu}_2^{+} = \mathbf{\mu}_2 + \mathbf{P}_2 (\mathbf{P}_1 + \mathbf{P}_2)^{-1} (\mathbf{\mu}_1 - \mathbf{\mu}_2).
    \end{align}
\end{subequations}
Just like for Eq. \ref{eq35:pfOfFusedCov}, we have to use the Woodbury matrix identity.
\begin{equation}\label{eq37:pfOfFusedMean}\begin{split}
    \mathbf{\mu}_1^{+} = \mathbf{\mu}_1 + \mathbf{P}_1 (\mathbf{P}_1 + \mathbf{P}_2)^{-1} (\mathbf{\mu}_2 - \mathbf{\mu}_1) \\
    = \mathbf{\mu}_1 + \mathbf{P}_1 (\mathbf{P}_1 + \mathbf{P}_2)^{-1} \mathbf{P}_1 [(\mathbf{P}_1)^{-1} \mathbf{\mu}_2 - (\mathbf{P}_1)^{-1} \mathbf{\mu}_1] \\
    = \mathbf{\mu}_1 + (\mathbf{P}_1 - \mathbf{P}_1^{+}) [(\mathbf{P}_1)^{-1} \mathbf{\mu}_2 - (\mathbf{P}_1)^{-1} \mathbf{\mu}_1] \\
    = \mathbf{\mu}_1 + (\mathbf{\mu}_2 - \mathbf{\mu}_1) - \mathbf{P}_1^{+} [(\mathbf{P}_1)^{-1} \mathbf{\mu}_2 - (\mathbf{P}_1)^{-1} \mathbf{\mu}_1] \\
    = \mathbf{\mu}_2 - \mathbf{P}_2^{+}[(\mathbf{P}_1)^{-1} \mathbf{\mu}_2 - (\mathbf{P}_1)^{-1} \mathbf{\mu}_1] \\
    = \mathbf{\mu}_2 + \mathbf{P}_2^{+} [(\mathbf{P}_1)^{-1} \mathbf{\mu}_1 - (\mathbf{P}_1)^{-1} \mathbf{\mu}_2] \\
    = \mathbf{\mu}_2 + \mathbf{P}_2^{+} (\mathbf{P}_1)^{-1} (\mathbf{\mu}_1 - \mathbf{\mu}_2) \\
    = \mathbf{\mu}_2 + \mathbf{P}_2 (\mathbf{P}_1 + \mathbf{P}_2)^{-1} (\mathbf{\mu}_1 - \mathbf{\mu}_2) = \mathbf{\mu}_2^{+}.
\end{split}\end{equation}
From Eqs. \ref{eq35:pfOfFusedCov} and \ref{eq37:pfOfFusedMean}, we have proven that the fused means and fused covariances for all GMM components will be the same, if we condition our estimate upon the other agent's \textit{a priori} mean. Next, we focus on the weight updates. 

Let us denote the weights of Agent 1's prior estimate as $\omega_{1,1}, \omega_{1,2}, ..., \omega_{1,N_1}$ and of Agent 2's prior estimate as $\omega_{2,1}, \omega_{2,2}, ..., \omega_{2,N_2}$. A formula for the weight update equation may then be given as follows:
\begin{equation} \label{eq38:weightUpdateDiff}
    \omega_{i_1,j_2}^{+} = \frac{\omega_{i_1} l_{i_1,j_2} \omega_{j_2}}{\sum_{m,n} \omega_{m_1} l_{m_1,n_2} \omega_{n_2}}.
\end{equation}
The subscripts $i_1$ and $j_2$ refer to the $i$-th and $j$-th GMM components of Agents 1 and 2, respectively. The likelihood function $l_{i_1,j_2}$ refers to the likelihood that the $i$-th GMM component mean of Agent 1 is the observation resulting from the $j$-th GMM component of Agent 2. A formula for this likelihood function is given below:
\begin{equation} \label{eq39:likelihoodP3}
    l_{i_1,j_2} = p_g(\mathbf{\mu}_{1,i} - \mathbf{\mu}_{2,j}; \mathbf{0}, \mathbf{P}_{xx,1,i} + \mathbf{P}_{xx,2,j}).
\end{equation}
We may interpret the likelihood as the innovation associated with Agent 1's use of Agent 2's GMM component means as a state estimate. If we switch the order of $i_1$ and $j_2$ in Eqs. \ref{eq38:weightUpdateDiff} and \ref{eq39:likelihoodP3}, then we will observe that the resulting weight update values for both equations are the same due to the same symmetry property of the Gaussian PDF. The fused estimate for both agents consists of $i_1j_2$ GMM components. This is because each mode of one agent's \textit{a priori} estimate is checked with the GMM modes of the other agent's \textit{a priori} estimate for consistency. Sometimes, this results in some fused GMM components vanishing and others standing out. 

The procedure for non-homogeneous estimate sensor fusion is summarized within Algorithm \ref{alg:nhe}. In the proceeding section, we demonstrate via simulation some GMM fusion cases for homogeneous and non-homogeneous \textit{a priori} estimates across a sensor network.

\begin{algorithm}
\caption{Consensus on Non-Homogeneous \textit{A Priori} GMMs}\label{alg:nhe}
\begin{algorithmic}[1]
    \State \parbox[t]{\dimexpr\linewidth-\algorithmicindent-1.5em}{\textbf{Input}: $[\mathbf{\mu}_{i_1}, \mathbf{\mu}_{j_2}, \mathbf{P}_{i_1}, \mathbf{P}_{j_2}, \omega_{i_1}, \omega_{j_2}]$  for all \textit{a priori} GMM components $i_1 \in \{1, 2, \cdots, N_1\}$ and $j_2 \in \{1, 2, \cdots, N_2\}$}
    \For{$i_1 \in \{1, 2, \cdots, N_1\}$}
        \For{$j_2 \in \{1, 2, \cdots, N_2\}$}
            \State \parbox[t]{\dimexpr\linewidth-\algorithmicindent-1.5em}{Perform a Kalman update on $\mathbf{\mu}_{i_1}$ using $\mathbf{\mu}_{j_2}$ as the observation using Eq. \ref{eq36:fusedMeans}.}
            \State \parbox[t]{\dimexpr\linewidth-\algorithmicindent-1.5em}{Perform a Kalman update on $\mathbf{P}_{i_1}$ using $\mathbf{P}_{j_2}$ as the observation using Eq. \ref{eq34:fusedCovs}.}
            \State \parbox[t]{\dimexpr\linewidth-\algorithmicindent-1.5em} {Compute means-as-observations likelihoods between all possible component fusions $(m_1, n_2)$ where $m_1 \in \{1, 2, \cdots, N_1\}$ and $n_2 \in \{1, 2, \cdots, N_2\}$ using Eq. \ref{eq39:likelihoodP3}.}
            \State \parbox[t]{\dimexpr\linewidth-\algorithmicindent-1.5em} {Compute the weight update $\omega_{i_1,j_2}^{+}$ using Eq. \ref{eq38:weightUpdateDiff}.}
        \EndFor
    \EndFor
\end{algorithmic}\end{algorithm}

\section{ILLUSTRATIVE EXAMPLES}

We demonstrate our sensor fusion techniques using two illustrative examples. The first example concerns a sensor network in which each agent has the same \textit{a priori} estimate, which is modeled as a GMM. In the second example, we demonstrate the fusion between two agents with different \textit{a priori} estimates, also modeled as GMMs.



\subsection{Homogeneous Prior Estimates}

For our first example, we consider a simple sensor network with three agents, with bidirectional communication between Agents 1 and 2 and between Agents 2 and 3. Each agent has the same, three-component GMM \textit{a priori} estimate of the target state, only one of which is consistent with the target truth. Furthermore, we utilize range (i.e. $l_2$-norm) as our measurement model. We illustrate our initial distribution and sensor network setup below.

\begin{figure}[thpb] 
      \centering

      \includegraphics[scale=0.5]{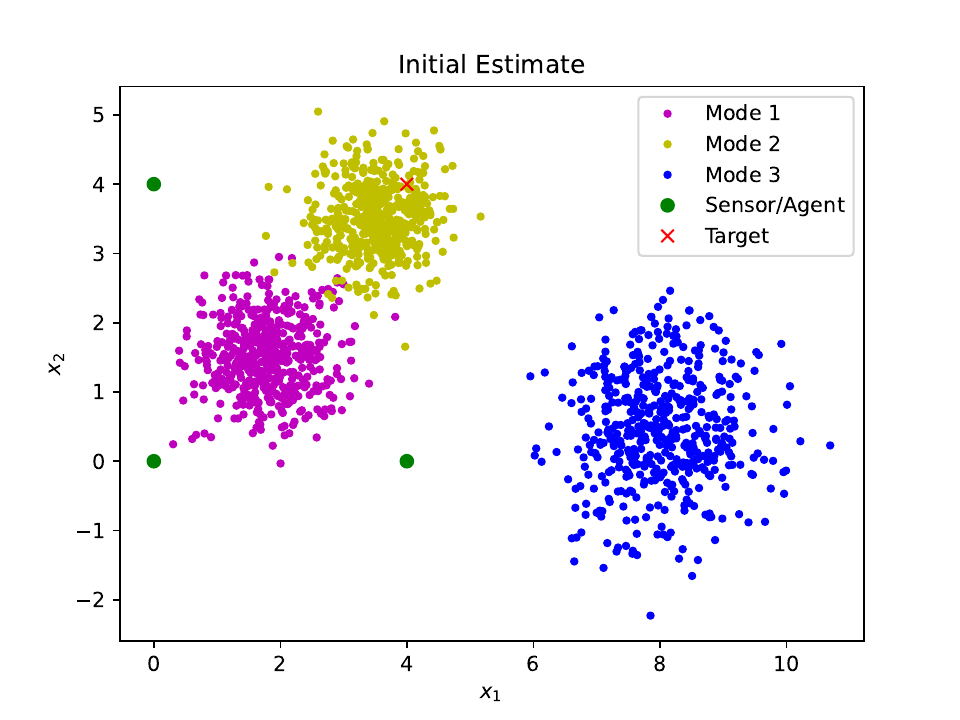}
      \caption{Distribution of each agent's \textit{a priori} estimate, locations of each sensor/agent (green), and target (red), which is located within the third GMM component}
      \label{fig1:initialP2}
\end{figure}

Since the target in Figure \ref{fig1:initialP2} lies squarely within the second GMM component, it is expected that the posterior estimate of all three agents is largely, if not entirely, the second GM component when updated by the observation. To compare our weights, we include a figure and table comparing the centralized and decentralized weight updates. The equation for the centralized weight update is given by Eq. \ref{eq24a:dUpdate}.

\begin{figure}[thpb] 
      \centering

      \begin{subfigure}{0.25\textwidth}
            \includegraphics[width=\linewidth]{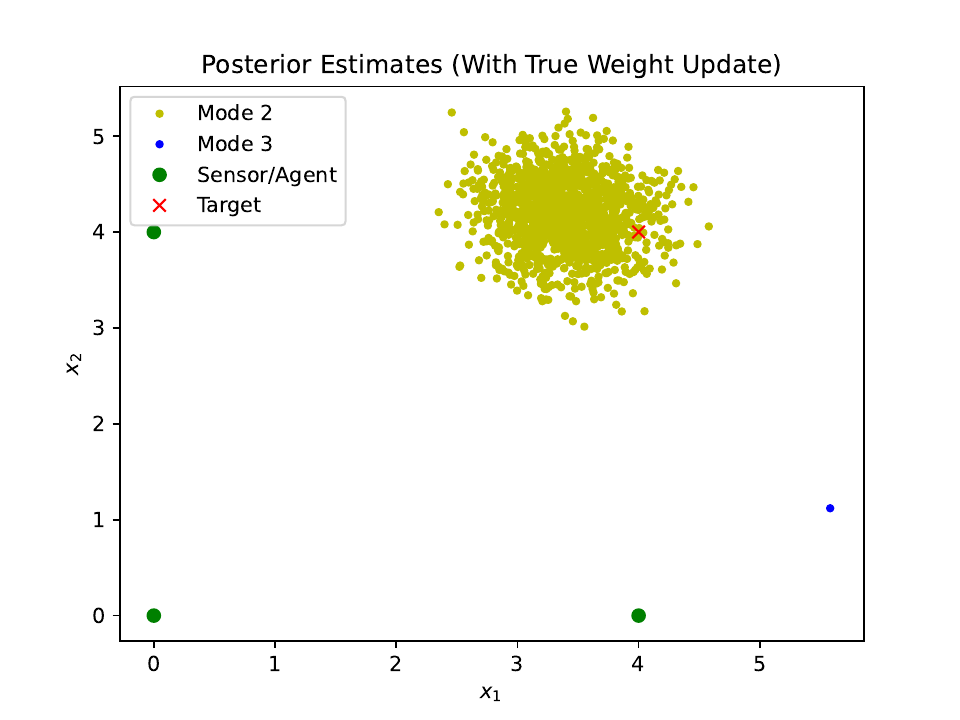}
            \caption{Centralized fusion estimate} \label{fig:2a}
      \end{subfigure}%
      \begin{subfigure}{0.25\textwidth}
            \includegraphics[width=\linewidth]{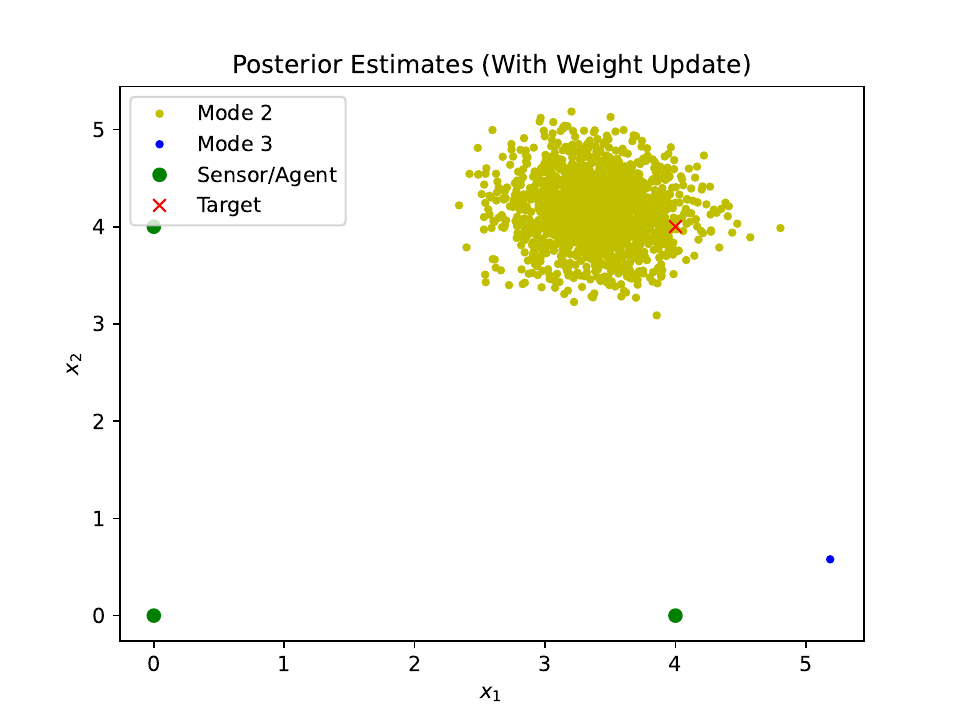}
            \caption{Distributed fusion estimate} \label{fig:2b}
      \end{subfigure}%
      \caption{Posterior estimates comparing a distribution drawn from a posterior GMM obtained using a centralized fusion method (left) and decentralized fusion method (right)}
      \label{fig2:posteriorEx1}
\end{figure}

\begin{table}[h]
\caption{Centralized and Decentralized Weight Updates}
\label{table:ex1}
\begin{center}
\begin{tabular}{|c||c||c||c|}
\hline
$i$ & 1 & 2 & 3\\
\hline
Prior $\omega_i$ & 0.333 & 0.333 & 0.334\\
\hline
Centralized Posterior $\omega_i$ & 0 & 0.9999 & 0.0001\\
\hline
Decentralized Posterior $\omega_i$ & 0 & 0.9999 & 0.0001\\
\hline
\end{tabular}
\end{center}
\end{table}
As expected, the second GMM component received the highest updated weight due to the target being contained within its $3\sigma$-interval. Furthermore, the new component weights with our distributed fusion method match those with a centralized fusion method. 


\subsection{Nonhomogeneous Prior Estimates}

\begin{figure}[thpb] 
      \centering

      \begin{subfigure}{0.25\textwidth}
            \includegraphics[width=\linewidth]{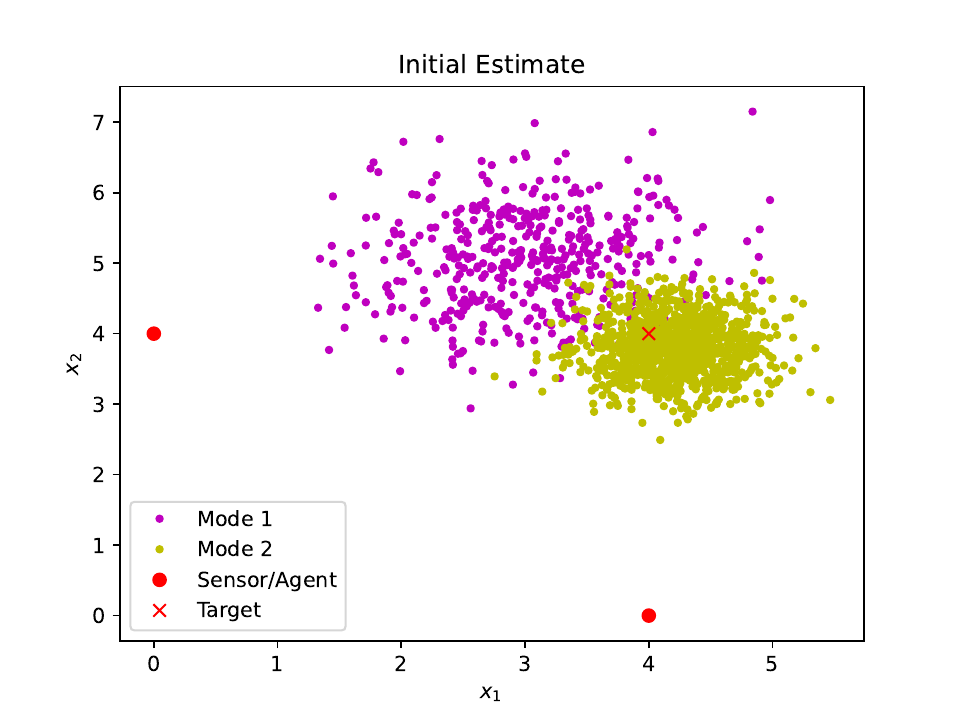}
            \caption{Agent 1's \textit{a priori} estimate} \label{fig:3a}
      \end{subfigure}%
      \begin{subfigure}{0.25\textwidth}
            \includegraphics[width=\linewidth]{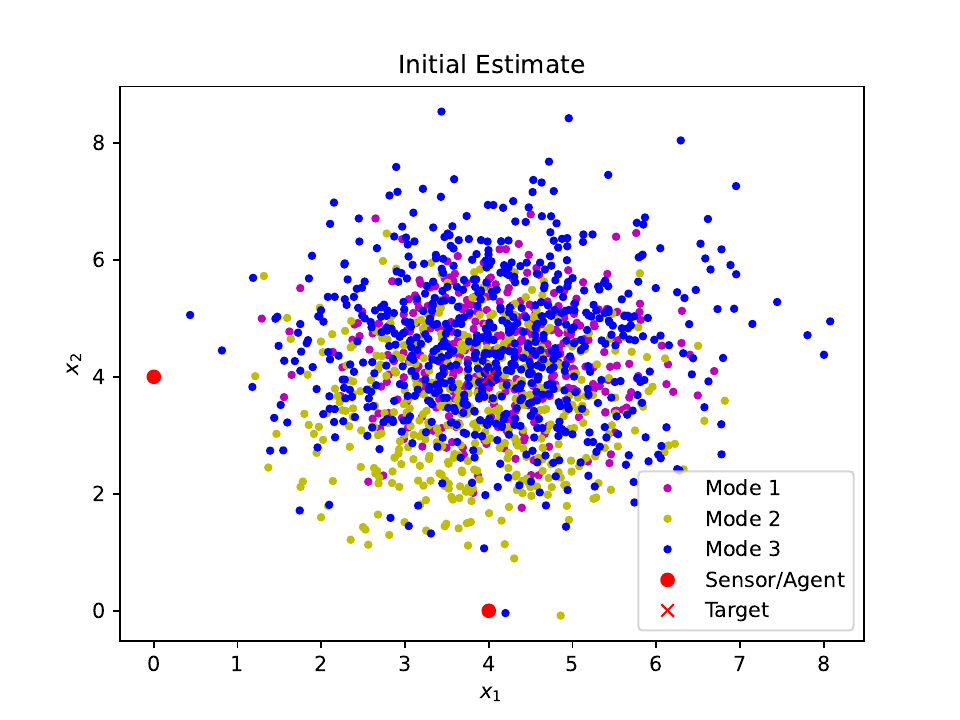}
            \caption{Agent 2's \textit{a priori} estimate} \label{fig:3b}
      \end{subfigure}%
      \caption{Particles drawn from \textit{A priori} GMM estimates from Agent 1 (left) and Agent 2 (right)}
      \label{fig3:priorEx2}
\end{figure}

In our second example, we remove the agent located at the origin in Figure \ref{fig1:initialP2} and connect the remaining two agents. Prior to that point, we shall assume there existed a disconnection which caused both sensors to obtain different \textit{a priori} estimates and even different numbers of GMM components. For this subsection, it is important to define two terms. The first is consistency, which refers to how much the overlap between particles drawn from one agent's prior estimate is to the particles drawn from the other agent's prior estimate. The other term is accuracy, which refers to whether there exists at least one GMM component from either agent's prior estimate which captures the target truth within its $3\sigma$-interval. We present an example in which all GMM components from each mode are both consistent and accurate, the initial distributions of which are given by Figure \ref{fig3:priorEx2}.

In the given example, the prior weights for Agent 1's GMM estimate were 0.3 and 0.7 for Modes 1 and 2, respectively, while the prior weights for Agent 2's tri-modal GMM estimate were 0.25, 0.3, and 0.45, respectively. The component weights, means, and covariances were chosen somewhat arbitrarily, so long as all GMM components for both agents' \textit{a priori} estimates were consistent and accurate. The following plots and table outline the resulting posterior distribution and association weights $\omega_{i_1, j_2}^{+}$. The association weights $\omega_{i_2, j_1}^{+}$ are simply a transpose of the Table \ref{table:ex2}'s weights.

\begin{figure}[thpb] 
      \centering

      \begin{subfigure}{0.25\textwidth}
            \includegraphics[width=\linewidth]{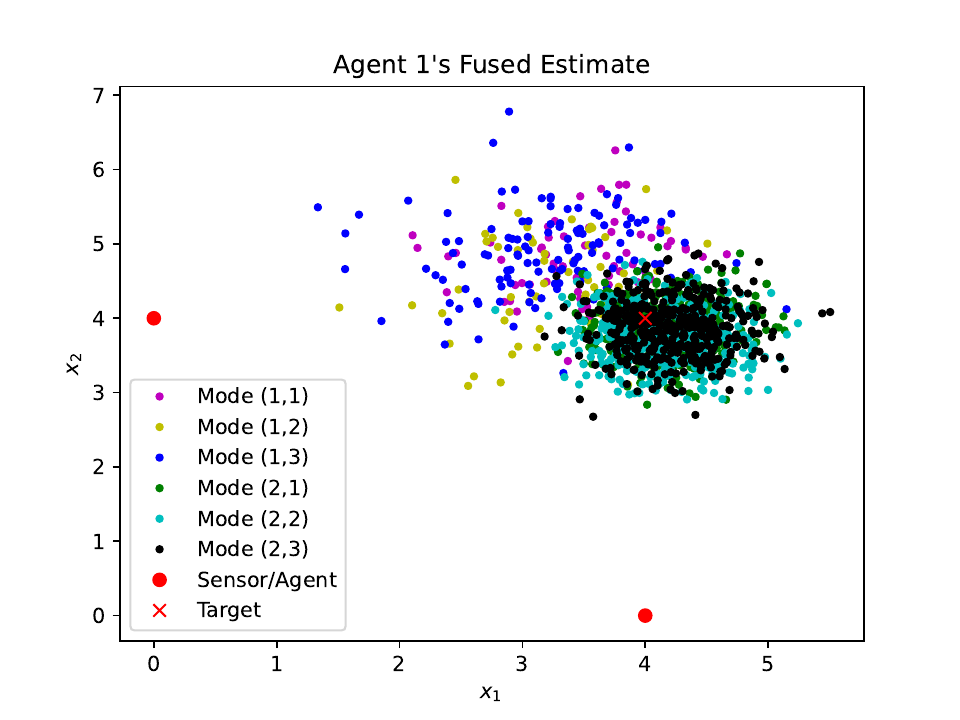}
            \caption{Agent 1's fused estimate} \label{fig:4a}
      \end{subfigure}%
      \begin{subfigure}{0.25\textwidth}
            \includegraphics[width=\linewidth]{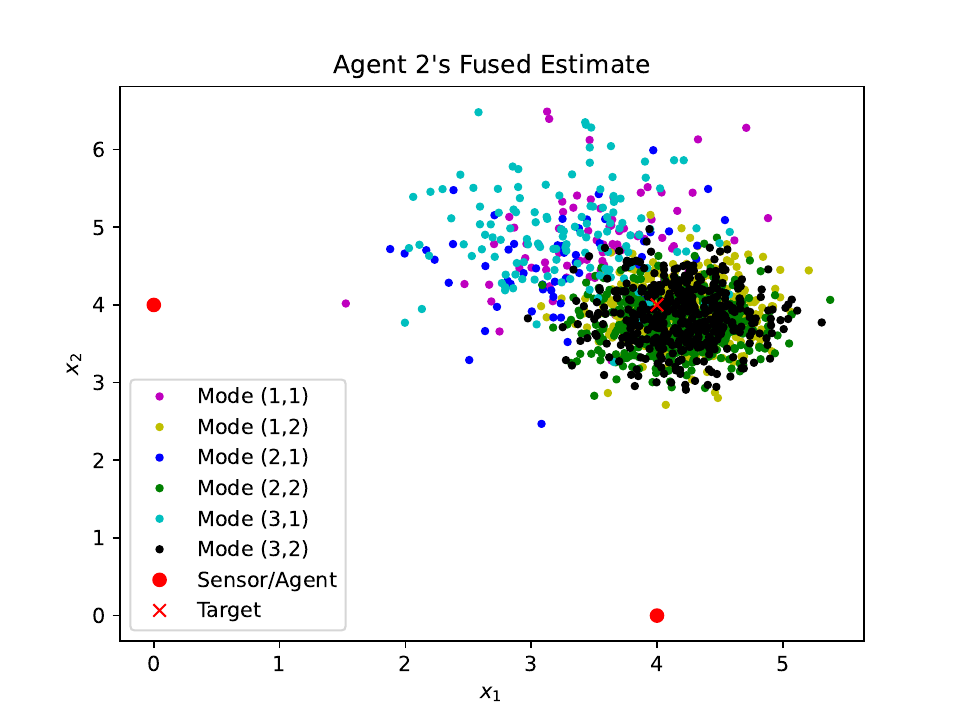}
            \caption{Agent 2's fused estimate} \label{fig:4b}
      \end{subfigure}%
      \caption{The fused estimates of Agents 1 and 2 are the same. Since Agent 1 had two \textit{a priori} GMM components and Agent 2 had three \textit{a priori} GMM components, the posterior distribution will contain six GMM components. Mode $(i_1, j_2)$ in \ref{fig:4a} has the same distribution as Mode $(j_2, i_1)$ in \ref{fig:4b}.}
      \label{fig4:postEx2}
\end{figure}

\begin{table}[h]
\caption{Decentralized Weight Updates}
\label{table:ex2}
\begin{center}
\begin{tabular}{|c||c||c||c|}
\hline
Posterior Weights $\omega_{i_1, j_2}^{+}$ & $i_1=1$ & $i_1=2$ & $i_1=3$\\
\hline
$j_2=1$ & 0.0489 & 0.0387 & 0.0866\\
\hline
$j_2=2$ & 0.2582 & 0.2774 & 0.2901\\
\hline
\end{tabular}
\end{center}
\end{table}

Since the consistency and accuracy definitions hold for all components across both agents, each component association has a non-insignificant (i.e. greater than 1\%) posterior weight, and the target truth is contained within each GMM component of the network's fused estimate. From this point, it is possible to use the DSE techniques outlined in Section \ref{sec:dse} to perform another update of this $i_1j_2$-component distribution. A key limitation to this fusion method is that it is scalable to only a two-agent/sensor system. If any more sensors are added to the network, this fusion method will encounter problems since the number of GMM mode associations grows exponentially with the number of agents, differing GMM components, and number of iterations. 

\section{CONCLUSIONS}

In this work, we studied how to use DSE techniques to fuse \textit{a priori} estimates modeled as GMMs across a decentralized network. We introduced two techniques which depended on two key assumptions: 1) whether each agent within the network had the same \textit{a priori} estimate or 2) whether each agent within the network had differing \textit{a priori} estimates. We demonstrated our methods using small-scale sensor network examples. 

A key improvement we wish to make for future studies is scalability. For the case of homogeneous \textit{a priori} estimates, we wish to study how to scale this distributed fusion method to networks of tens, even hundreds, of sensors/agents and more GMM components. While our current fusion method for the case of non-homogeneous prior estimates is scalable for only two sensors/agents, we want to better model real-world sensor fusion scenarios, in which several sensors/agents have differing \textit{a priori} estimates and may experience frequent disconnections or jamming. While we only show sensor fusion across a single time step in this work, we also wish to explore how well our sensor fusion methods scale with time, including when the network undergoes random disruptions.

\addtolength{\textheight}{-12cm}   





\section{ACKNOWLEDGMENT}

The authors acknowledge and are grateful for funding from AFRL Space Vehicles Chief Scientist Innovative Research Program.



\end{document}